# Benchmarking LiDAR Sensors for Development and Evaluation of Automotive Perception

Fredrik Schalling
KTH Royal Institute of Technology
Stockholm, Sweden
schalli@kth.se

Sebastian Ljungberg
KTH Royal Institute of Technology
Stockholm, Sweden
sljungb@kth.se

Naveen Mohan
KTH Royal Institute of Technology
Stockholm, Sweden
naveenm@kth.se

*Abstract*—Environment perception and representation are some of the most critical tasks in automated driving. To meet the stringent needs of safety standards such as ISO 26262 there is a need for efficient quantitative evaluation of the perceived information. However, to use typical methods of evaluation, such as comparing using annotated data, is not scalable due to the manual effort involved. There is thus a need to automate the process of data annotation. This paper focuses on the LiDAR sensor and aims to identify the limitations of the sensor and provides a methodology to generate annotated data of a measurable quality. The limitations with the sensor are analysed in a Systematic Literature Review on available academic texts and refined by unstructured interviews with experts. The main contributions are 1) the SLR with related interviews to identify LiDAR sensor limitations and 2) the associated methodology which allows us to generate world representations.

## I. INTRODUCTION

The development of automated driving is advancing rapidly and several *Original Equipment Manufacturers (OEMs)* are currently testing prototypes with a high *level* (as defined by SAE J3016 [1]) of automation on public roads. In essence, the decisions made for longitudinal and lateral control of a vehicle, are made on the basis of a world representation created by using a multitude of sensors. To ensure that the decisions are correct, a common validation technique used is pre-recorded annotated data colloquially known as *Ground Truth (GT)* [2] or performing a qualitative evaluation by manual intervention.

While using GT and qualitative evaluations provide a solution to the problem of verification and validation, to demonstrate that no errors have crept in from the development process, the solution must be analysed from the perspective of functional safety standards such as ISO 26262 [3] and SOTIF [4]. ISO 26262 was designed to provide a framework on addressing functional safety in safety-critical automotive functions. Following the standard entails demonstrating evidence and reasoning in a cogent *safety case* that the safety requirements are complete and satisfied by the evidence. While ISO 26262 was not created particularly for automated vehicles, some studies have been made that show that meeting the stringent requirements of ISO 26262 could require evidence of several billions of kilometers driven [5].

Currently, there exists no general framework for reasoning about the performance of perception systems [6] and existing methods require either GT or human intervention. With standards as ISO 26262 in mind and the billions of kilometers that need to be tested, it is inevitable to automate i.e. avoid human intervention for as much of the process of evaluation of perception systems as possible.

This paper aims to investigate the possibility of creating a GT equivalent named *Pseudo Ground Truth (PGT)* using information gained exclusively from a LiDAR sensor. Furthermore we aim to discuss which precautions is needed in the development of perception systems if a proposed algorithm is used to counteract *limitations* with the LiDAR technology is used. We believe that in using an automatically generated PGT greatly increases the amount of data available than a typically manually generated GT, thereby helping OEMs step closer to meeting the requirements for billions of miles of driving, while providing greater data variation from real-world scenarios.

To understand where PGT can be used as a valid substitute for GT, this paper addresses the following research question:

**Which precautions need to be taken in the development of perception systems for automated vehicles when only using a high precision LiDAR sensor as a substitute for GT information?**

The question is intended to be answered through a literature study to identify limitations of the LiDAR technology to which precautions needs to be taken. The literature study is followed up with a proposal of how to tackle some of the identified limitations and a proposal of how to find when the PGT could be used as GT. A more elaborate definition of the study is made in Section II and information found in the literature study is described in Section III-A. Lastly, an experimental study is proposed in Section IV. The result of the study and its limitations are discussed in Section V and finally the conclusions of this study are presented in Section VI.

## II. METHOD

Fig 1 shows the discrete steps that we have undertaken in this work and point to the section numbers where they can be found.

A deductive research approach is chosen as framework of the study, where the first part contains steps to identify the limitations of the LiDAR sensor from existing academic literature and industrial experts. The literature is examined in a systematic way using the guidelines for a *Systematic Literature Review* (SLR) by Kitchenham et al. [7] with the aim to identify limitations with a LiDAR-based perception system in an automotive application. The review protocol contains the steps followed enabling the repeatability making the the bias in the literature review easier to evaluate [7], [8]. The findings from this SLR are used as a basis for discussion with experts in the field to identify the most critical limitations.

The second part is an experimental study, where the most critical limitations of the LiDAR technology is addressed through an algorithm that implements a PGT. The experiments are used to evaluate possible solutions to reduce some of the stated limitations of a LiDAR in the literature review, hence reducing the amount of precautions that are needed to be taken if the proposed verification system is deployed in the development of perception systems. To follow up the simulation validating tests in a real-world setup is used. Lastly, the study and its contribution is concluded.

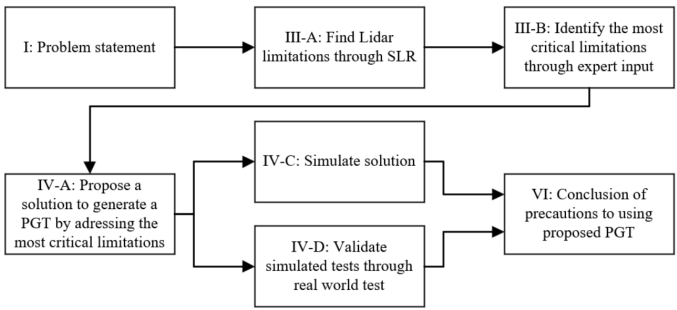

Fig. 1. The discrete sequences in the methodology

## III. IDENTIFYING LIMITATIONS FROM STATE OF THE ART AND PRACTICE

This section details the identification of the limitations of LiDAR sensors, performed using the SLR and input from industry experts.

### A. Literature review

A SLR is performed in two areas namely: environmental and technical limitations of using the LiDAR technology in an automotive application. The study is carried out in accordance to Kitchenham et al. [7]. The review planning has been iterated to find a set of keywords, inclusion- and exclusion criteria that generate papers that are relevant to this work. The final version is presented in this paper.

**RQ focus:** To identify and characterise limitations of the LiDAR technology in the context of automotive applications.
**RQ:** Under what conditions does a LiDAR sensor risk to perform significantly worse than expected?
**Context:** The scenarios have to either (1) directly relate to an automotive application or (2) are reasonably likely to be found within an automotive application.

**Source List:** The Google Scholar database has been used as the sole source, as it indexes all databases of interest to us such as IEEE, Springer etc.
**Inclusion and Exclusion criteria:** As LiDAR is a relatively new technology for the automotive industry, older publications will not be included. Furthermore, only papers that address perception related to an automotive application will be included since the use case can differ between industries. This reasoning results in the following inclusion and exclusion criteria.

**Inclusion criteria:**
- I1 The publication is published the year 2015 or later.
- I2 The publication is among the top 50 search results.

**Exclusion criteria:**
- E1 The publication is not directly related to automotive perception and includes a limitation of the using the LiDAR technology in perception.
- E2 The publication is not available in English and in full format online.

The material in the literature review was collected in March 2019 through the search strings, shown in Table I, are based on the key words stated in the review planning.

TABLE I
SEARCH STRINGS

| |
|---|
| lidar AND performance AND automotive AND limitations AND problem AND weather |
| – |
| (lidar OR "laser radar") AND performance AND automotive AND environment AND test |

The SLR resulted in 91 unique publications of which exclusion criteria E1 and E2 led to 58 respective 9 exclusions. The final resulting 24 publications were read thoroughly and the limitations found are specified in Table II, categorised into three distinct categories; Obstructing (Ob), Attenuating/Noise (AN) and Other (Ot). Furthermore, the citations is classified depending on if the limitation found is either mentioned where it is not the primary context (Mention) and if it is a primary source and main content of its paper (Experimental).

*1) Obstructing conditions:* Among the 24 identified publications seen in Table II, 8 limitations are either mentioned or shown in an experiment to give an obstructing impact. In this category the effects of targets consisting of a different material or having different surfaces are also included. Generally, this category depicts limitations with a predictable or controllable impact. For example, road dirt accumulated over a long time on the sensor cover as in [10] could be controlled through simple maintenance of the vehicle or, as shown in [13] the impact of a wet surface is predictable and has around 10 % reduced reflectivly.

*2) Attenuating/Noise conditions:* Among the 24 identified publications seen in Table II, 17 papers are found to address attenuating or noise induction. 11 of the found limitations are categorised as experimental. Most common are scenarios of precipitation, which includes both refraction, when a particle reflects enough effect back to the sensor to be detected as a hit, and an unordered scattering of the beams. These scenarios generally induce disturbances with an irregular behaviour including both an attenuating effect and generation of noise. Rain for example is shown to, as expected, be dependent on the wavelength used in the LiDAR. As showed in [27] the wavelength is set to 905 nm in 95 % of the LiDAR sensors on the market but a more expensive technology based on 1550 nm laser, shows a more robust performance in rainy conditions.

TABLE II
LIMITATIONS TO THE LiDAR SENSOR TECHNOLOGY

| Ind. | Limitation | Mention | Experimental | Cat. |
|---|---|---|---|---|
| 1 | Road dirt on sensor cover | [9] | [10] | Ob |
| 2 | First detection close object | | [11] | Ob |
| 3 | Material/surfaces | [12] | [13] [14] [11] | Ob |
| 4 | Wet roadway causes road spray | | [15] | AN |
| 5 | Rain | [16] [17] [18] [19] [20] [21] [9] | [22] [13] [23] [24] [25] | AN |
| 6 | Fog/Mist/Haze | [16] [17] [18] [19] [25] [20] [21] [9] | [26] [27] [28] [23] | AN |
| 7 | Snow | [16] [17] [18] [19] [20] [9] | [26] [23] | AN |
| 8 | Dust | [17] [23] | | AN |
| 9 | Wavelength related | | [19] | AN |
| 10 | Sunlight | [29] [18] | [15] [30] | AN |
| 11 | Temperature | [31] | | Ot |
| 12 | Vibrations | [31] | | Ot |
| 13 | Interference | | [32] | Ot |
| 14 | Remote attacks (imitating signal) | | [33] | Ot |

*3) Other conditions:* 8 instances of very specific limitating scenarios are found. This category includes limitations as vibrations induced by uneveness of the road or uneven mixing of fuel and oxygen combustion which over time causes deteriorated performance of the sensor [31]. Another limitation is when the number of active LiDAR sensors rise in traffic the risk of interference between them increase which is assessed in [32].

The most common limitations to the LiDAR technology have been identified as limitations 5-7 which are all different forms of precipitation. This category generally gives either a reduced line of sight and/or high uncertainties of readings which increases with range from the sensor. This category also induces random noise which must be countered actively to create a robust representation of the environment. The changes in properties between material or wetness of the surface, is also an important factor since the behavioural change could be large though generally predictable.

*B. Expert interview*

To rate the importance of the findings of the SLR in the automotive application input is gathered from a professional at Scania CV. The input validated the extensiveness and confirmed the emphasis on the importance of the limitations with unpredictable noise in precipitation. The discussions performed so far have been informal for the purposes of early feedback. A more thorough structured interview is planned in the near future.

IV. EXPERIMENTAL DESIGN

An experimental study is proposed to evaluate the effects of the findings in the SLR and with that, also the limitations of using a LiDAR sensor to generate PGT. The limitations 5-7 in Table II will be addressed with a method to reduce their impact on the generation of PGT, followed up with simulated tests to quantify under which precautions the PGT is a valid substitute for the GT. The performance metrics for tests in a simulated world are proposed as correlation measures between the generated PGT versus GT. This evaluation should thus help evaluate from under which conditions the PGT is a valid substitution for GT, i.e. minimize the error between GT and PGT. Lastly, real-world tests are proposed as a validation of simulated results to confirm that the simulation results are applicable for real sensors and environments.

*A. Solution to limitations*

We propose adding a filter as shown effective in [34] to remove precipitation from sensor scans, and then addressing the issue of sparsity with accumulation over time as shown robust in [35]. The accumulation of point clouds will be used to create the PGT through merging the sensor readings prior to mapping them into a global world representation. Since we deal with a final global representation, we are less susceptible to temporary changes in the environment and need only logs from the real-world along with the sensor characteristics to generate PGT. The real-world tests will be performed in an instrumented area of the Scania testing tracks where GT can be obtained for comparison relatively easily.

*B. Inputs to experiments*

To ensure that the experiments accurately portray automotive applications, technologies and scenarios, the following LiDAR products, mapping algorithm and KPIs are considered.

LiDAR devices are a laser-based ranging system that is based on time-of-flight on the reflected pulse, and are able to measure a distance towards an object. Long range LiDAR sensors has a maximum range of 100 to 300 m but works best from closer than 50 m. They can have a field of view angle up to 360°. The current state of the art LiDAR has a cycle time between 20 to 50 ms. Table III below shows the evaluated LiDAR devices.

Grid maps are used as the world representation of the LiDAR sensor readings. The grid map discretizes the world into cells. Based on the sensor data, every cell is determined by a probabilistic function that decides whether it is occupied or free. Grid maps were chosen as the world representation due to their use in similar projects such as [2] to enable

TABLE III
LiDAR PRODUCT MODELS ON THE MARKET

| Model | Range | FoV - horizontal | Accuracy (distance, degrees) | Cycle time (FoV) |
|---|---|---|---|---|
| Quanergy M8-1 | 150 m | 360° | 5 cm, 0.03° | 33 ms |
| Ibeo LUX | 200 m | 110° | 10 cm, 0.125° | 20 ms |
| Continental SRL1 | 10 m | 27° | 10 cm, - | 10 ms |
| Velodyne HDL-64ES2 | 120 m | 360° | 2 cm, 0.09° | 50 ms |
| Velodyne Alpha Puck | 300 m | 360° | 3 cm, 0.11° | 50 ms |
| Ouster OS-2 | 250 m | 45° | -, 0.175° | 50 ms |

the comparison of results and the ease of data conversions as explained in Section IV.

The study has to be done by doing a comparison based on the geographically overlapping PGT and GT cells. Quality measures as Map Score, Occupied Cells ratio [2] and Pearsons correlation coefficient are to be used.

*C. Simulation*

To be able to assess the solution, test in simulations will be done, the setup is shown in Figure 2,

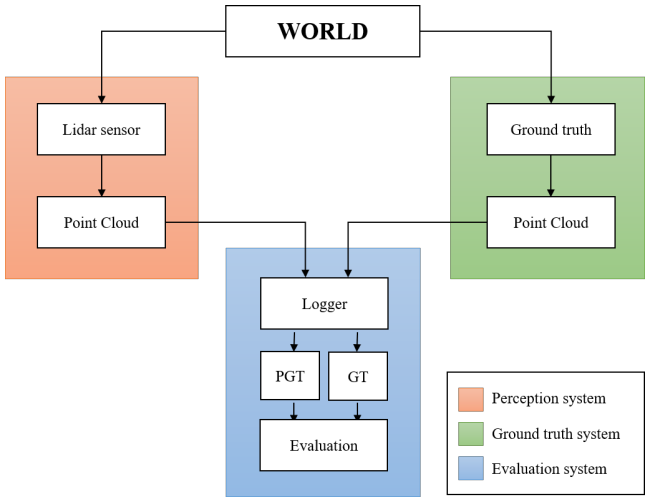

Fig. 2. Flowchart of information in a simulated environment

where the orange section is the perception system which creates PGT. The green section is created by a GT sensor in the simulation with the same range as the LiDAR model. The blue section receives all data in point clouds with a known position for every time step in the Logger for a specific test case. The Logger store information about the sensor model and the GT in a frame for every sequence. Two grid maps for every state will be the output, the PGT and GT. The simulating environment is setup in accordance to the solution in [36]. The setup of the simulation environment is done and the test cases and evaluation will be performed in April 2019.

*D. Real-world tests*

To find where the simulation of both sensor readings and noise models are correct real-world tests are proposed. These tests will be used to find whether the same tendencies are found in reality as shown in the simulations. To facilitate this a RTK enhanced GPS is to be used to generate GT in conjunction with known points in the environment instrumented with GPS transceivers. In a setup like this there is a possibility to set up identical test cases in the simulation and real-world to evaluate the accuracy of the simulated results. The authors have access to a test track and tests will be performed in late 2019.

V. DISCUSSION

A threat to validity of this work is the focus on literature from 2015, because of the assumption that the LiDAR technology is evolving at a high pace. This could mean that potentially useful articles from before 2015 have been excluded. Another threat to completeness in our literature review is that e.g. in Table II, the column "mention" lists limitations of the LiDAR sensors that are discussed in particular papers. There is a risk that this could bias our findings towards what is generally accepted from the research community. We have not followed through and critically examined where the mentioned limitations were obtained from. A similar bias could be induced by the inclusion of words such as environment or weather, i.e. some prior knowledge of limitations with the LiDAR technology which may have limited some potentially useful papers. To counter these weaknesses both an industry expert is consulted and primary sources are found to the deemed "most critical" limitations of the LiDAR technology. While the SLR has been initiated, we intend to use the snowball sampling method to mitigate these weaknesses by expanding our included publication list.

Although, using occupancy grids as world representation is suitable for evaluation of an environment with irregular shapes and enables certain efficient correlational KPIs, it has a few disadvantages. Firstly, it will have discretization errors by design which has to be weighted towards the potential sparsity based on the sensor resolution. Also, the choice of occupancy grids is not ideal for tests in a dynamic environment. The occupancy grid does not by default contain temporal information and would also add difficulties in the accumulation of data into a global representation. So, in an environment containing more dynamic changes than our use-cases, the choices of world representation could be unsatisfactory. This could potentially be addressed by choosing a more suitable world representation.

In the evaluation, we are expecting the KPIs do degrade between ideal condition compared to in a scenario of an identified limitation and also in a similar manner in both simulation and real-world test. When the validity is verified the simulation is to be used to quantify when the PGT consists of enough information to be used as GT. When having the quantified performance of the PGT this method is proposed to be used in the development of the perception system. Enabling this feedback loop should allow for a more agile development process not bounded by what annotated data is available, but only the access to testing scenarios.

It could be argued that some limitations to the LiDAR sensor could be solved by using other sensor technologies. However, our work is limited to charting the limitations of LiDAR sensors and provides an essential input to finding solutions. Potential solutions to these limitations e.g. detecting objects using other sensors etc., although the next logical step, remains out of scope.

## VI. Conclusion/Future work

The limitations of building a perception system relying exclusively on the LiDAR technology are found through an SLR and presented. From these limitations, the most critical limitations, for our work, have been identified as precipitation based scenarios. The limitations have been assessed and an algorithm is proposed to reduce their impact through a data accumulation process, generating PGT.

The proposed algorithm to generate PGT is then evaluated in simulation to check if it is a valid substitution for GT under a set of precautions (derived from the limitations). The algorithm is designed to enable an automated and quantitative evaluation of perception systems. Today the first three steps in Figure 1 have been performed. Furthermore, the simulation setup in [36], will be augmented with the the analytical implementations described in Section IV-C within the near future to perform a quantitative evaluation under multiple scenario.

## VII. Acknowledgment

The authors gratefully acknowledge the following projects and agencies for financial support: H2020 - ECSEL – AutoDrive (Grant Agreement number: 737469), and H2020 - ECSEL PRYSTINE (Grant agreement number 783190). We also would like to thank our primary collaborator at Scania CV, Hjalmar Lundin for his support and initiation of the project.